\documentclass[aps,prb,twocolumn,superscriptaddress,floatfix,letter]{revtex4-1}
\usepackage{epsfig,amsmath,amssymb,color}
\begin{document}

\title{Chebyshev matrix product state approach for time evolution}
\author{Jad C. Halimeh} 
\affiliation{Physics Department and Arnold Sommerfeld Center for Theoretical Physics, Ludwig-Maximilians-Universit\"at M\"unchen, 80333 M\"unchen, Germany}
\author{Fabian Kolley} 
\affiliation{Physics Department and Arnold Sommerfeld Center for Theoretical Physics, Ludwig-Maximilians-Universit\"at M\"unchen, 80333 M\"unchen, Germany}
\author{Ian P. McCulloch} 
\affiliation{ARC Centre for Engineered Quantum Systems, School of Mathematics and Physics, The University of Queensland, St Lucia, Queensland 4072, Australia }

\date{\today}

\begin{abstract}
We present and test a new algorithm for time-evolving quantum many-body systems initially proposed by Holzner {\em et al.} [Phys.\ Rev.\ B {\bf 83}, 195115 (2011)]. The approach is based on merging the matrix product state (MPS) formalism with the method of expanding the time-evolution operator in Chebyshev polynomials. We calculate time-dependent observables of a system of hardcore bosons quenched under the Bose-Hubbard Hamiltonian on a one-dimensional lattice. We compare the new algorithm to more standard methods using the MPS architecture. We find that the Chebyshev method gives numerically exact results for small times. However, the reachable times are smaller than the ones obtained with the other state-of-the-art methods. We further extend the new method using a spectral-decomposition-based projective scheme that utilizes an effective bandwidth significantly smaller than the full bandwidth, leading to longer evolution times than the non-projective method and more efficient information storage, data compression, and less computational effort.
\end{abstract}

\maketitle
\section{Introduction}
 Besides being of central interest in the field, achieving large accessible times in the time evolution of strongly-correlated quantum many-body systems with existing numerical methods has proven to be a daunting task in any spatial dimension and particularly for global quenches. Experiments in quantum many-body physics in the last years have evolved in such a manner that local control over degrees of freedom has become more feasible\cite{Greiner,Fukuhara,Ronzheimer,Bloch,Cheneau,Trotzky,Trotzky2,Simon} and in which quantum magnetism, spin dynamics, and relaxation dynamics have been explored. Additionally, along this experimental work a whole body of theoretical investigations has arisen that relies on various analytical and numerical methods to describe the dynamics of these experiments.
 
The time-dependent Schr\"odinger equation
\begin{equation}
\text{i}\frac{d}{dt}|\psi(t)\rangle = \hat{H}|\psi(t)\rangle
\end{equation}
for a generic time-independent Hamiltonian $\hat{H}$ and initial state $|\psi_{0}\rangle=|\psi(0)\rangle$ is formally solved by the time evolution operator
\begin{equation}
U(t) = \exp( -\text{i} \hat{H} t),
\end{equation}
where the reduced Planck constant $\hbar$ is set to $1$. The time-evolved quantum state for arbitrary times is then given by
\begin{equation}\label{eq:tevolve}
|\psi(t)\rangle = U(t)|\psi_{0}\rangle = \exp(-\text{i}\hat{H}t)|\psi_{0}\rangle.
\end{equation}
However, for a many-body quantum system, the dimension of the Hilbert space grows exponentially with the number of constituents in the system under consideration, making it impossible to calculate the matrix exponential in Eq.~\eqref{eq:tevolve} exactly and, therefore, approximate methods are required.

The purpose of this work is to discuss achievable evolution times for complex quantum many-body systems such as global quenches relevant to the current experimental efforts in the field. As the errors encountered in experiments are usually larger than those in numerical calculations, we are not interested in an increase in accuracy.

One method that has proven extremely useful is $t$-DMRG\cite{White,Schollwoeck,White1992,Uli007}, which is based on the description of the quantum state in terms of matrix product states (MPS)~\cite{Ostlund1995, Dukelsky1998,Uli,Verstraete2006,Gobert,Feiguin,Ripoll}
\begin{equation}
|\psi\rangle = \sum_{\mathbf{\{\sigma\}}}c_{\mathbf{\sigma}}|\mathbf{\sigma}\rangle = \sum_{\{\mathbf{\sigma}\}}A^{\sigma_{1}}\dots A^{\sigma_{N}}|\mathbf{\sigma}\rangle,
\end{equation}
where $\mathbf{\sigma} = \{\sigma_{1}\dots \sigma_{N} \}$ is the computational basis, $A^{\sigma_{1}}$ and $A^{\sigma_{N}}$ are $D$-dimensional row and column vectors, respectively, and $A^{\sigma_{i}}$ ($i=2,\dots,N-1$) is a $D\times D$ matrix. Theoretically, every quantum state can be represented by an MPS if infinite matrix dimensions are allowed~\cite{Vidal1}. The practical relevance of such a state description lies in the fact that one can often very well approximate the exact quantum state by an MPS with finite matrix dimension. From this perspective, MPS presents a class of states that compress exact many-body quantum states such that the number of coefficients needed to describe the state scales linearly in the number of constituents as opposed to the exponential scaling in the exact representation. Furthermore, the approximation made in the compression step is well understood\cite{Verstraete2006} and can be controlled by the matrix dimension $D$.
 

With the help of MPS, several methods have been developed to calculate the time evolution of one-dimensional many-body quantum systems\cite{Uli}. The earliest methods utilize the Trotter\cite{Trotter,Suzuki} decomposition of the time-evolution operator. Later approaches approximate the matrix exponential in the Krylov\cite{Krylov} subspace. Both methods have been successfully applied to a series of different physical problems.

Nevertheless, the times reachable with current methods are still very limited making the development of new methods still a very important endeavor. The limitation of evolution times accessible with MPS-based methods is closely related to the amount of entanglement in the quantum state. The maximal entanglement between two subsystems describable by an MPS is given by the logarithm of the matrix dimension $D$. On the other hand, it has been shown that the entanglement after a quantum quench grows typically linearly in time~\cite{Calabrese} leading to an exponentially-growing matrix dimension, which is required in order to keep the error fixed. 

In this work, we test a new method for calculating the time evolution of one-dimensional quantum many-body systems as it was proposed in Ref.~\onlinecite{Holzner} by Holzner {\em et al.} We attempt to merge MPS with the method of approximating the time-evolution operator in terms of Chebyshev polynomials. The procedure of expanding the time-evolution operator in terms of Chebyshev polynomials is general and requires in principle solely a matrix-vector multiplication. The MPS approach together with the representation of the Hamiltonian as a matrix product operator provides an efficient way to perform these operations in the quantum many-body framework. A related approach based on Chebyshev polynomials has recently been successfully applied in the frequency domain to obtain an efficient impurity solver for the dynamical mean-field theory (DMFT) algorithm\cite{Wolf,Ganahl,Tiegel} and for calculating spectral functions\cite{Wolf2} and Green's functions\cite{Schmitteckert}.

For real-time dynamics in MPS, however, the approach has not been tested so far. In this paper, we test the new method (dubbed $t$-CheMPS) for a non-trivial system of hardcore bosons which evolve in time under the Bose-Hubbard Hamiltonian in a one-dimensional lattice. We show that time-dependent observables can be calculated numerically exactly with the $t$-CheMPS method up to a certain time beyond which exponentially growing errors become dominant. The time reachable is given by the amount of entanglement in the $n$-th Chebyshev vector and can be slightly increased by making use of a projection procedure onto the energy range where the initial state has finite nonzero spectral weight. We compare our results to the time-evolution methods based on the Trotter\cite{Trotter,Suzuki} expansion and Krylov\cite{Krylov} approximation of the time-evolution operator. We find that for the problem considered in this work the Trotter-based method reaches the longest times, followed by the method based on the Krylov approximation. 

This paper is structured as follows: Section \ref{standard} gives a brief overview of the standard state-of-the-art methods in time evolution within the MPS context. Section \ref{tCheMPS} discusses the $t$-CheMPS method and its workings. Section \ref{ptCheMPS} presents an extension of the latter, namely, \emph{projective} $t$-CheMPS based on the spectral decomposition of the initial state. Section \ref{model} discusses the Bose-Hubbard-model global quench used for the simulations in this paper. the results of which are documented in Section \ref{results}. The paper concludes with Section \ref{conclusion}. 

\section{Standard time-evolution methods in MPS}\label{standard}
\subsection {Krylov time-evolution}
Instead of treating Schr\"{o}dinger's equation as a differential equation, one considers, for time-independent Hamiltonians, the time-evolution operator $\exp({-\text{i}\hat{H}t})$. This sets the nontrivial task of evaluating an exponential of matrices\cite{Moler,Uli007}. One of the most efficient methods is the so-called Krylov subspace approximation\cite{Krylov,Uli,Uli007}, where one realizes that our interest lies in $\exp({-\text{i}\hat{H}t})|\psi\rangle$ rather than $\exp({-\text{i}\hat{H}t})$. In DMRG $\hat{H}|\psi\rangle$ is available efficiently, and this can be utilized through forming the Krylov subspace by successive Gram-Schmidt orthonormalization of the set $\{|\psi\rangle,-\text{i}\hat{H}t|\psi\rangle,(-\text{i}\hat{H}t)^2|\psi\rangle,\cdots\}$, where $|\psi\rangle$ is assumed to be normalized here. Here, $-\text{i}\hat{H}t$ is approximated regarding its extreme eigenvalues by $VTV^T$, where $V$ is the matrix containing the $n$ Krylov vectors thus obtained from the Gram-Schmidt decomposition and $T$ is an $n\times n$ tridiagonal matrix. This approximation is up to a very good precision even for relatively small numbers of Krylov vectors\cite{Uli007}. Thereafter, the exponential is given by the first column of $V\exp T$, where the latter exponential is now much easier to calculate.

\subsection{Suzuki-Trotter time-evolution}
Another prominent and very efficient method for evaluating the above matrix exponential is the (Suzuki-)Trotter decomposition\cite{Trotter,Suzuki,Uli,Uli007}. This method is mainly useful for Hamiltonians with nearest-neighbor interactions. In the case of a one-dimensional chain, the Hamiltonian $\hat{H}=\hat{H}_1+\hat{H}_2$ is divided into odd- and even-bond terms, $\hat{H}_1$ and $\hat{H}_2$, respectively, where $\hat{H}_1=\sum_{i=1}^{N/2}\hat{h}_{2i-1}$ and $\hat{H}_2=\sum_{i=1}^{N/2}\hat{h}_{2i}$. Here, $\hat{h}_i$ is the local Hamiltonian linking sites $i$ and $i+1$, and $N$ is the total number of sites on the lattice. $[\hat{H}_1,\hat{H}_2]\neq 0$ as neighboring local Hamiltonians do not commute in general, but all the terms in $\hat{H}_1$ and $\hat{H}_2$ commute. As such, the first-order Trotter decomposition of the infinitesimal time-evolution operator is

\begin{equation}
e^{-\text{i}\hat{H}\Delta t}=e^{-\text{i}\hat{H}_1\Delta t}e^{-\text{i}\hat{H}_2\Delta t}+\mathcal{O}(\Delta t^2).
\end{equation}

\noindent Moreover, the second-order Trotter decomposition reads

\begin{equation}
e^{-\text{i}\hat{H}\Delta t}=e^{-\text{i}\hat{H}_1\Delta t/2}e^{-\text{i}\hat{H}_2\Delta t}e^{-\text{i}\hat{H}_1\Delta t/2}+\mathcal{O}(\Delta t^3).
\end{equation}

\noindent One can go for yet higher orders and conclude that an $n^{th}$-order Trotter decomposition will yield over a time step $\Delta t$ an error of the order of $(\Delta t)^{n+1}$. As one requires $t/\Delta t$ time steps in order to reach an evolution time $t$, the error grows at worst linearly\cite{Uli007} in time $t$, and therefore, the resulting error is bound by an expression of the order of $(\Delta t)^nt$. For the purposes of this study, it turns out that second-order Trotter decomposition is optimal.

Time-dependent DMRG ($t$-DMRG) uses adaptive Hilbert spaces that follow the state $|\psi(t)\rangle$ being optimally approximated, and was first proposed independently in the works of Daley, Kollath, Schollw\"{o}ck, and Vidal\cite{Daley} and White and Feiguin\cite{White}, based on the time-evolving block-decimation (TEBD) algorithm\cite{Vidal1,Vidal2} for the classical simulation of the time evolution of weakly-entangled quantum states. Shortly afterwards, Schmitteckert~\cite{Schmitteckert1} published on nonequilibrium electron transport in interacting one-dimensional spinless Fermi systems using $t$-DMRG.

\section{$t$-C\MakeLowercase{he}MPS} \label{tCheMPS}
In this section, we review a recipe for time evolution using the Chebyshev matrix product state approach, namely $t$-CheMPS, as it was proposed in Ref.~\onlinecite{Holzner} by Holzner \emph{et al.}  As such, a brief expos\'e on Chebyshev polynomials is in order.

Chebyshev polynomials of the first kind, $T_n(x);\;n\in\mathbb{N}$ are given by the recursive relations

\begin{equation}
  T_n(x) = \left\{
  \begin{array}{l l}
    1 & \quad \text{for $n=0$},\\
    x & \quad \text{for $n=1$},\\
    2xT_{n-1}(x)-T_{n-2}(x) & \quad \text{for $n>1$}.\\
  \end{array} \right.
\end{equation}

\noindent A useful non-recursive expression for the Chebyshev polynomials is 

\begin{equation}
T_n(x)=\cos(n\arccos x).
\end{equation}

\noindent Moreover, they form an orthonormal set of polynomials on the interval $x\in[-1,1]$ with respect to the weighted scalar product

\begin{equation}
 \langle T_{n},T_{m}\rangle = \int_{-1}^{1} \frac{dx}{\pi\sqrt{1-x^2}}T_{n}(x)T_{m}(x), 
\end{equation}

\noindent and are divergent in the region $x\in(-\infty,-1)\cup(1,\infty)$.  

Chebyshev polynomials have been extensively studied in the mathematics and engineering literature\cite{Abramowitz,Rivlin,Boyd,Mason,Weisse}.

\subsection{Chebyshev expansion in the time domain}
\begin{figure}[]
\includegraphics[scale=0.8]{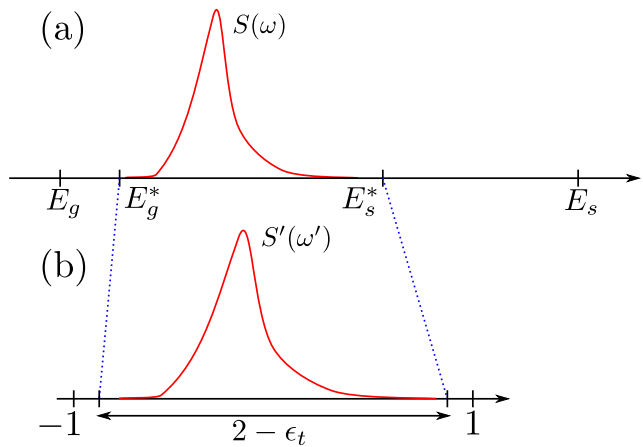}
\caption{(Color online)
(a) The spectral decomposition $S(\omega)$ of $|\psi_0\rangle$ relative to $\hat{H}$ has nonzero weight in the region $[E_g^*,E_s^*]$ where the \emph{effective}  bandwidth $W^*=E_s^*-E_g^*$ is significantly smaller than the full many-body bandwidth $W=E_s-E_g$. (b) In the Chebyshev expansion approach for time evolution, it may be advantageous to rescale $\hat{H}$ by mapping the effective bandwidth from $[E_g^*,E_s^*]$ to $[-W',W']$ where $W'=1-\frac{\epsilon_t}{2}$ with $\epsilon_t=0.025$ being a safety factor\cite{Holzner}.
}
\label{fig:spectral}
\end{figure}
We consider a system in which a Hamiltonian $\hat{H}$ acts on an initial state $|\psi_0\rangle$, thus propagating its time evolution.  The full many-body bandwidth of $\hat{H}$ is $W=E_s-E_g$, where $E_g$ ($E_s$) is the groundstate (sky\-state) energy of $\hat{H}$.  In many cases, this bandwidth is far larger than the \emph{effective} bandwidth $W^*=E_s^*-E_g^*$ that one can determine from the spectral function of $|\psi_0\rangle$ relative to $\hat{H}$.  As illustrated in Fig. \ref{fig:spectral}, the spectral function of $|\psi_0\rangle$ relative to $\hat{H}$ has nonzero weight mainly over $[E_g^*,E_s^*]$. Since the Chebyshev polynomials of first kind are divergent outside of the region $x\in[-1,1]$, and knowing that these polynomials will be functions of a Hamiltonian, this effective bandwidth is rescaled to $[-W',W']$ where $W'=1-\frac{\epsilon_t}{2}$ and $\epsilon_t$ is a safety factor to guarantee that the domain of the Chebyshev polynomials will remain within $\mathcal{I}=[-1,1]$.  In our numerical simulations, $\epsilon_t$ has been set to $0.025$.  This rescaling, when applied to the original Hamiltonian $\hat{H}$ will lead to a rescaled Hamiltonian $\hat{H}'$ where

\begin{equation}\label{eq:rescale}
\hat{H}'=\frac{\hat{H}-b}{a},
\end{equation}

\noindent with $a=W^*/(2-\epsilon_t)$ and $b=(E_g^*+E_s^*)/2$.  Now one can express the Chebyshev polynomials of the first kind in terms of this rescaled Hamiltonian.

Several constructions of Chebyshev approximations\cite{Weisse} can be found for a given function $f(x)|_{x\in\mathcal{I}}$, with the one most suited for our purposes being

\begin{equation}\label{eq:construct}
f(x)=\frac{1}{\pi\sqrt{1-x^2}}\left[\mu_0+2\sum_{n=1}^{\infty}\mu_nT_n(x)\right],
\end{equation}

\noindent where the Chebyshev moments $\mu_n$ are given by

\begin{equation}\label{eq:chebymoments}
\mu_n=\int_{-1}^1f(x)T_n(x)dx.
\end{equation}

\noindent An order of $N$ approximation $f_N(x)$ of $f(x)$ is possible if one has access to the first $N$ terms ($0\leq n\leq N-1$), and thus it follows that

\begin{equation}\label{eq:chebyrep}
f_N(x)=\frac{1}{\pi\sqrt{1-x^2}}\left[\mu_0+2\sum_{n=1}^{N-1}\mu_nT_n(x)\right].
\end{equation}

\noindent The time-evolution operator can be expressed as ($\hbar=1$)

\begin{equation}\label{eq:TEO_Chebyshev}
\hat{U}(t)=e^{-\text{i}\hat{H}t}=\int_{-1}^1d\omega 'e^{-\text{i}[a(\omega '+W')+E_g^{*}]t}\delta(\omega'-\hat{H}'),
\end{equation}

\noindent and upon expressing the $\delta$-function term therein as per Eq. (\ref{eq:chebyrep}), one obtains

\begin{equation}\label{eq:TEO}
\hat{U}_N(t)=e^{-\text{i}(E_g^{*}+aW')t}\sum_{n=0}^{N-1}\phi_n(t)T_n(\hat{H'}),
\end{equation}

\noindent with $\phi_0(t)=c_0(t)$ and $\phi_{n>0}(t)=2c_n(t)$, where

\begin{equation}\label{eq:cn}
c_n(t)=\int_{-1}^1\frac{e^{-\text{i}at\omega'}T_n(\omega')}{\pi\sqrt{1-\omega'^2}}d\omega'=(-\text{i})^nJ_n(at),
\end{equation}

\noindent and $J_n(at)$ is the Bessel function of the first kind of order $n$.

It is to be noted here that in $t$-CheMPS, as will be elucidated later, one does not have to calculate the actual wavefunction $|\psi_N(t)\rangle=\hat{U}_N(t)|\psi_0\rangle$ at a Chebyshev order $N$ in order to determine the time evolution of some observable.


\subsection{Recipe for time evolution of initial state $|\psi_0\rangle$}
Here, we provide the steps needed to time-evolve an initial state $|\psi_0\rangle$ under a Hamiltonian $\hat{H}$ using the $t$-CheMPS method. As an initialization step, we calculate the groundstate $|g\rangle$ and the skystate $|s\rangle$ of $\hat{H}$, noting that the skystate of $\hat{H}$ is nothing but the groundstate of $-\hat{H}$.  This allows us to determine the bandwidth $W$ of $\hat{H}$, from which we can make a specific choice for $W^*$ using the spectral-decomposition technique highlighted in the next section.  Then we can determine $a$ and $b$ and rescale $\hat{H}$ to $\hat{H'}$ as per Eq. (\ref{eq:rescale}). The first Chebyshev vector $|t_0\rangle$ is set to the initial state $|\psi_0\rangle$, while the second Chebyshev vector is given by $|t_1\rangle=\hat{H'}|t_0\rangle$.  Thereon, any Chebyshev vector $|t_{n\geq2}\rangle$ is obtained via the recursive relation

\begin{equation}\label{eq:recurrence}
|t_n\rangle=2\hat{H'}|t_{n-1}\rangle-|t_{n-2}\rangle.
\end{equation}

\noindent This recurrence relation can be implemented using the compression or fitting procedure\cite{Uli,Holzner}.  This procedure finds an MPS representation for $|t_n\rangle$ by variationally minimizing the \emph{fitting error}\cite{Holzner}

\begin{equation}\label{eq:fitting_error}
\Delta_{\text{fit}}=\big|\big||t_n\rangle-(2\hat{H'}|t_{n-1}\rangle-|t_{n-2}\rangle)\big|\big|^2.
\end{equation} 

\noindent This procedure of \emph{recurrence fitting} effects variational minimization through a sequence of sweeps back and forth along the chain that proceed until the state being optimized becomes stationary.  Calling the state $|t_n\rangle$ and $|t'_n\rangle$ after and before a fitting sweep, it becomes stationary once the term

\begin{equation}\label{eq:deltac}
\Delta_c=\left|1-\frac{\langle t_n|t'_n\rangle}{\big|\big||t_n\rangle\big|\big|\cdot\big|\big||t'_n\rangle\big|\big|}\right|
\end{equation}

\noindent drops below a specified \emph{fitting convergence threshold}\cite{Holzner}, which we have determined to suffice when set to $10^{-6}$ for our purposes.

\subsection{Energy truncation}

The DMRG truncation step in the recursive-fitting procedure where $\hat{H'}$ is applied onto $|t_{n-1}\rangle$ in order to calculate $|t_n\rangle$ (see Eq. (\ref{eq:recurrence})) is not performed in the eigenbasis of $\hat{H'}$, and, as such, high-energy components can be possibly passed on to subsequent recursion steps, leading to divergences in higher-order Chebyshev vectors\cite{Holzner}. This is remedied via energy truncation sweeps that occur locally at each site through building the corresponding Krylov subspace, proceeding with the energy truncation at the site, and completing it before moving on to the next site. As DMRG truncation occurs in the recurrence-fitting procedure, no such further truncation is carried out here. The energy eigenbasis of $\hat{H'}$, where the energy truncation is to be performed, is not possible to access in full, and thus a Krylov subspace of dimension $d_K$ is constructed at each site. Then, a method such as Arnoldi's algorithm is utilized to calculate the extreme eigenvalues of $\hat{H'}$ that are bigger than an energy truncation error bound per time step $\varepsilon$ in magnitude, where one can set $\varepsilon=1.0$, and focus on the proper value of $W^*$ based on the spectral decomposition of the initial state $|\psi_0\rangle$ with respect to $\hat{H'}$.  This is due to the fact that whatever value of $W^*$ one picks, the range of effective eigenenergies $[E_g^*,E_s^*]$ will be rescaled to $[-W',W']$, and in the Chebyshev context, the maximum and minimum energies must be no larger than $\varepsilon$ in magnitude.  Further details on this method can be found in Ref. \onlinecite{Holzner}.

\subsection{Computing the time evolution of an observable}\label{timeconstraint}
Consider that we wish to compute the time evolution 

\begin{equation}\label{eq:TE1}
\langle\hat{O}_j\rangle(t)=\langle\psi(t)|\hat{O}_j|\psi(t)\rangle
\end{equation} 

\noindent of some observable $\hat{O}$ at a given site $j$ on the chain.  We represent the time-evolved state $|\psi(t)\rangle$ in terms of the Chebyshev representation of order $N$ of the time-evolution operator of Eq.~\eqref{eq:TEO} on the initial state $|\psi_0\rangle$:

\begin{equation}\label{eq:obs1}
|\psi(t)\rangle=e^{-\text{i}(E_g^*+aW')t}\sum_{n=0}^{N-1}T_n(\hat{H'})\phi_n(t)|\psi_0\rangle,
\end{equation}

\noindent Noticing that $|\psi_0\rangle=|t_0\rangle$ and that $T_n(\hat{H'})|t_0\rangle=|t_n\rangle$, Eq. (\ref{eq:obs1}) becomes

\begin{equation}\label{eq:obs2}
|\psi(t)\rangle=e^{-\text{i}(E_g^*+aW')t}\sum_{n=0}^{N-1}\phi_n(t)|t_n\rangle.
\end{equation}

\noindent Plugging Eq. (\ref{eq:obs2}) into Eq. (\ref{eq:TE1}), we get

\begin{equation}\label{eq:TE2}
\langle\hat{O}_j\rangle(t)=\sum_{n,m=0}^{N-1}\phi^{\ast}_m(t)\phi_n(t)\langle t_m|\hat{O}_j|t_n\rangle.
\end{equation} 

As already mentioned, in $t$-CheMPS one is never obligated to calculate the actual wavefunction $|\psi(t)\rangle$ itself in order to calculate a certain observable using Eq. (\ref{eq:TE2}). Furthermore, the coefficents $\phi_{n}(t) = (-i)^{n}J_{n}(at)$ ($n>0$) decay rapidly with $n$ for $n>at$. It is therefore possible to define a maximaum time for a given number of Chebyshev moments $N$ such that the neglected weight in terms of the coefficients is smaller than a certain threshold. We define $t_{\text{max}}$ as the largest $t$ such that 

\begin{equation}\label{eq:fabian}
\sum_{n=N}^{\infty}\phi_{n}^{*}(t)\phi_{n}(t) < 10^{-3},
\end{equation}

\noindent which is justified because the moments $\langle t_{m} | \hat O_j|t_{n}\rangle$ decay quickly with $|n-m|$ [see Fig.~\ref{fig:cheb_matrix2}]. In practice we determine $t_{\text{max}}$ by calculating $\sum_{n=N}^{N_{\text{max}}}\phi_{n}^{*}(t)\phi_{n}(t) < 10^{-3}$, with $N_{\text{max}} = 500$ for which we have $\phi_{N_{\text{max}}}(t)<10^{-100}$ in the relevant time range or $\phi_{N_{\text{max}}}(t)=0$ for all practical purposes.

\section{Projective $t$-C\MakeLowercase{he}MPS}\label{ptCheMPS}

We wish to find a way to calculate the effective bandwidth of a wavefunction $|\psi\left(t\right)\rangle$ at a time $t$.  The reason behind this is that in $t$-DMRG one uses the \emph{full} bandwidth while time-evolving the wavefunction and that leads to smaller evolution times that can be reached numerically.  Using a smaller \emph{effective} bandwidth may lead to larger numerically-accessible evolution times.

Let the full many-body bandwidth of the model be $W = E_s- E_g$. Suppose that the initial state $|\psi_0\rangle = |\psi(t=0) \rangle$ has
spectral support on a limited frequency interval $[E_g^*, E_s^*]$, of width  $W^* = E_s^*- E_g^*$, where $E_s^*<E_s$ and $E_g^*>E_g$. Then, 
it would be possible to do the time evolution with $t$-CheMPS by 
rescaling this effective bandwidth, rather than the full bandwidth, onto the interval $[-1, 1]$.  Thus, it is of interest to explore the spectral decomposition
of the initial state $|\psi_0\rangle$, and of its time-evolved
version, $|\psi(t)\rangle$. We now discuss how this can be done,
focussing first on $|\psi_0\rangle$, and thereafter generalizing
the discussion to  $|\psi(t)\rangle$ in the Appendix. For ease of notation, in this Section and the Appendix, $\hat{H}$ shall denote the rescaled Hamiltonian of our system with bandwidth $W=E_s-E_g$, and $\omega\in [-1,1]$.

\subsection*{Spectral decomposition of initial state $|\psi_0\rangle$}\label{sec:spectral_decomp}
The spectral decomposition of $|\psi_0\rangle$ is

\begin{equation}\label{eq:S} S(\omega) = \langle\psi_0|\delta(\omega-\hat{H})|\psi_0\rangle.\end{equation}

\noindent A Chebychev expansion of the $\delta$-function of order $N$ has the form:

\begin{multline} \label{eq:delta} \delta_N(\omega-\hat{H})= \\
\frac{1}{\pi\sqrt{1-\omega^2}}\left[g_0+2\sum\limits_{n=1}^{N-1}g_n T_n(\hat{H})T_n(\omega)\right],
\end{multline}

\noindent where the coefficient $g_n$ is a Jackson damping coefficient defined as

\begin{equation} \label{eq:Jackson} g_n=\frac{(N-n+1)\cos\frac{\pi n}{N+1}+\sin\frac{\pi n}{N+1}\cot\frac{\pi}{N+1}}{N+1}. \end{equation}

\noindent We introduce $\theta_n$ such that

\begin{equation}
  \theta_n = \left\{
  \begin{array}{l l}
    g_0 & \quad \text{if $n=0$ },\\
   2 g_n & \quad \text{if $n>0$}.\\
  \end{array} \right.
\end{equation}

\noindent This allows us to write Eq. (\ref{eq:delta}) as

\begin{equation} \label{eq:deltaN} \delta_N(\omega-\hat{H})=\frac{1}{\pi\sqrt{1-\omega^2}}\sum\limits_{n=0}^{N-1}\theta_n T_n(\hat{H})T_n(\omega). \end{equation}

\noindent Now we calculate $S(\omega)$ using Eq. (\ref{eq:deltaN}) and noting that our initial wavefunction $|\psi_0\rangle$ equals the first Chebyshev vector $|t_0\rangle$ and that $|t_n\rangle=T_n(\hat{H})|t_0\rangle$:

\begin{align} \label{eq:SN} \nonumber S(\omega) & =   \langle\psi_0|\delta_N(\omega-\hat{H})|\psi_0\rangle \\&= \frac{1}{\pi\sqrt{1-\omega^2}}\sum\limits_{n=0}^{N-1}\theta_nT_n(\omega)\langle t_0|t_n\rangle,
\end{align}

Hence, all we have to do to calculate $S(\omega)$ is to calculate the moments $\mu_n=\langle t_0|t_n\rangle$.  To achieve a specified spectral resolution of, say, $\Delta$, we have to use an expansion order of $N_\Delta = 2 W/ \Delta$. Moreover, we provide in the Appendix a derivation in terms of the Chebyshev moments of the spectral decomposition of the time-evolved wavefunction $|\psi(t>0)\rangle$, which can be used as a numerical-fidelity check.

\section{Global-Quench Test Model} \label{model}
\noindent For the comparison we wish to carry out between the Suzuki-Trotter decomposition, the Krylov approxi\-mation and the $t$-CheMPS methods, we consider a benchmark test model: a strong global quench in the Bose-Hubbard model (BHM) on a bosonic lattice at half filling with odd-site unity filling for different values of the on-site interaction strength $U$. Global quenches happen when an initial state undergoes a time evolution due to a new Hamiltonian for which the initial state has an extensively different energy as for the original Hamiltonian whose groundstate it was. 

\begin{figure}[]
\includegraphics[scale=0.45]{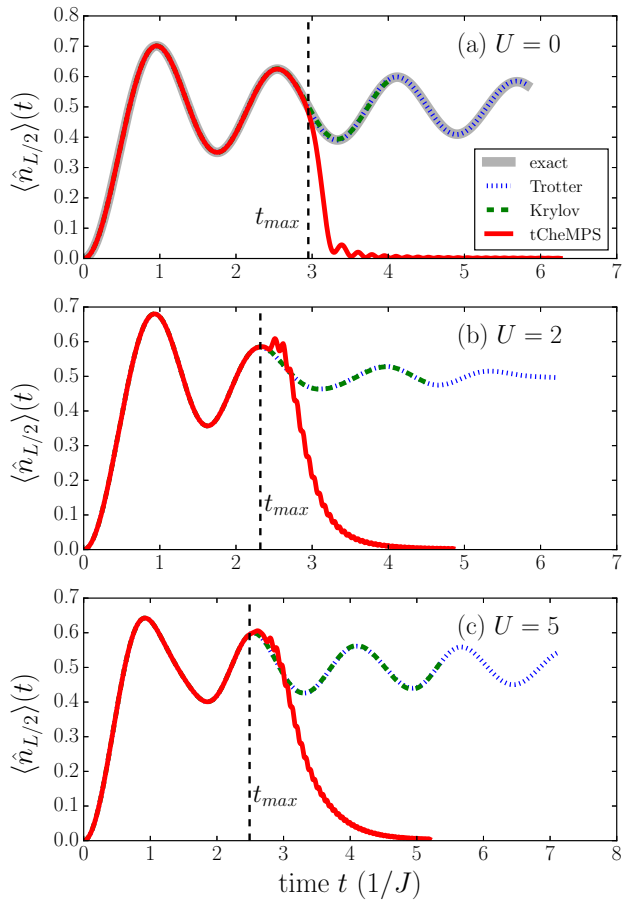}
\caption{(Color online)
The time evolution for the particle density at site $L/2$ after a global quench with (a) $U=0$, (b) $U=2$, and (c) $U=5$, each obtained with the Trotter (blue dotted), Krylov (green dashed) and $t$-CheMPS (red solid) methods. For $U=0$ the exact time evolution given by Eq.~\eqref{eq:exact_u0} is shown in light grey. All the methods give numerically exact results for short times. The time reached by the $t$-CheMPS method is given by $t_{\text{max}}$ (see Eq.~\eqref{eq:fabian}). After this time threshold the error quickly increases. In all cases the Trotter method reaches the longest times followed by the Krylov method.
}
\label{fig:convergence}
\end{figure}

We consider an initial state

\begin{equation}\label{eq:initial_state}
|\psi_0\rangle=|\psi(0)\rangle=\prod_{i=1}^{L/2}\hat{b}_{2i-1}^{\dagger}|0\rangle
\end{equation}
that is a bosonic lattice of size $L=32$ in which every odd site has a single boson and every even site holds zero occupancy. It can be thought of as the groundstate of some suitable Hamiltonian. The system is globally quenched to the Bose-Hubbard Hamiltonian 

\begin{equation}\label{eq:Hamiltonian}
\hat{H}=-J\sum_{i=1}^{L-1}\left(\hat{b}_i^{\dagger}\hat{b}_{i+1}+h.c.\right)+\frac{U}{2}\sum_{i=1}^{L}\hat{n}_i(\hat{n}_i-1),
\end{equation}

\noindent where $J$ and $U$ are the hopping and interaction terms of the Bose-Hubbard model.  This global quench has already been studied using $t$-DMRG \cite{Cramer,Cramer1,Flesch}. We consider different values of the on-site interaction strength, including the analytically solvable case of $U=0$.

At $U=0$, the Hamiltonian in Eq. (\ref{eq:Hamiltonian}) reduces to 

\begin{equation}\label{eq:Hamiltonian1}
\hat{H}=-J\sum_{i=1}^{L-1}\left(\hat{b}_i^{\dagger}\hat{b}_{i+1}+h.c.\right).
\end{equation}

\begin{figure*}[]
\includegraphics[width=6.3in]{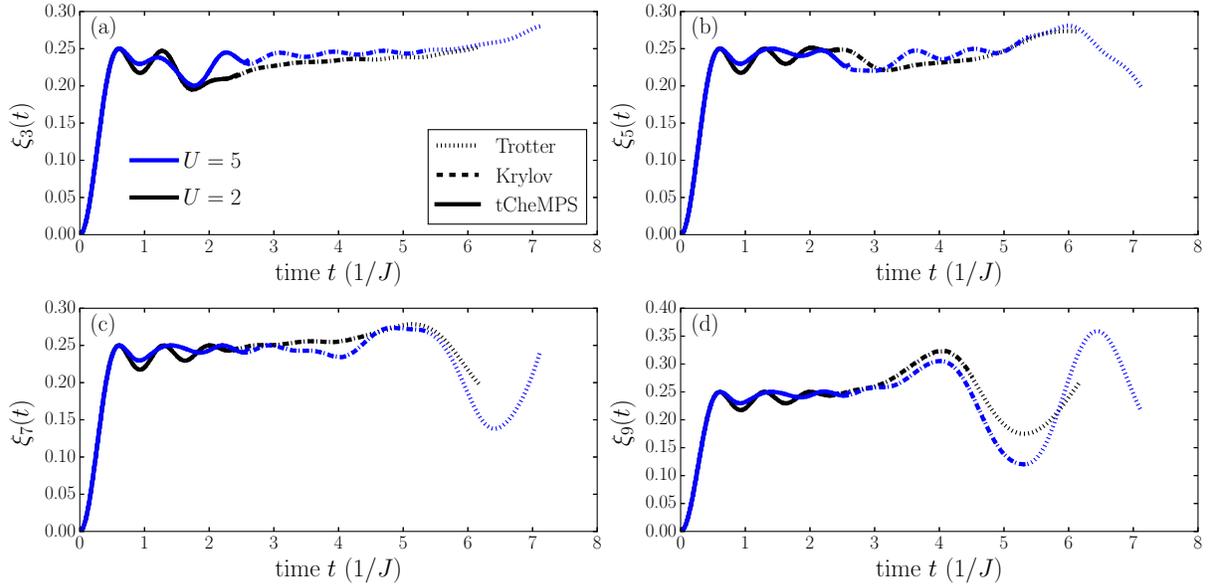}
\caption{(Color online)
Density-density correlations $\xi_j(t)=\langle\psi(t)|\hat{n}_{L/2-j}\hat{n}_{L/2+j-1}|\psi(t)\rangle$ for $U=2$ (black) and $U=5$ (blue) are shown for different half distances $j$: (a) $j=3$, (b) $j=5$, (c) $j=7$, and (d) $j=9$. All the correlators are obtained by the Trotter (dotted), Krylov (dashed), and $t$-CheMPS (solid) methods. All the methods give the same results up to the corresponding reachable times.
}
\label{fig:denden}
\end{figure*}

In the case of non-interacting bosons ($U=0$), scattering is not the physical mechanism behind local relaxation. Instead, the time-dependent contributions to the reduced density operator of the regarded subsystem consist of quickly oscillating phases that average out under sufficient conditions leading to a relaxation of the density operator\cite{Flesch}. In the current case, excitations start propagating from all sites with a finite speed throughout the duration of the time evolution spreading the information about the initial conditions more and more over the entire system. The incommensurate mixing of these excitations then can lead to a state that appears to be locally perfectly relaxed. This case leads to an exact analytical solution covered in Ref. \onlinecite{Flesch} by a Fourier transformation of the ladder operators involved. In the Heisenberg picture the time evolution of the ladder operators reads 

\begin{align}
\hat{b}_i(t)&=\frac{1}{L}\sum_{k}\sum_{l=1}^{L}e^{-\text{i}k(l-i)}e^{2\text{i}J\cos(k)}\hat{b}_l(0), \\
\hat{b}_i^{\dagger}(t)&=\frac{1}{L}\sum_{k}\sum_{l=1}^{L}e^{\text{i}k(l-i)}e^{-2\text{i}J\cos(k)}\hat{b}_l^{\dagger}(0),
\end{align}

\noindent where $k=\frac{2\pi}{L}l$, where $l=1,2,\ldots,L$. As $\hat{n}_i(t)=\hat{b}_i^{\dagger}(t)\hat{b}_i(t)$, one obtains

\begin{multline}\label{eq:exact_u0}
\langle\hat{n}_i\rangle(t) = \frac{1}{2}\left(1+\frac{1}{L}\sum_{q=0}^{L-1}(-1)^{i+1}e^{-4\text{i}Jt\cos\left(\frac{2\pi}{L}q\right)}\right) \\ \xrightarrow{L\rightarrow\infty}\frac{1}{2}\left(1+(-1)^{i+1}J_0(4Jt)\right).
\end{multline}

\section{Results and discussion}\label{results}

\begin{figure}[]
\includegraphics[scale=0.45]{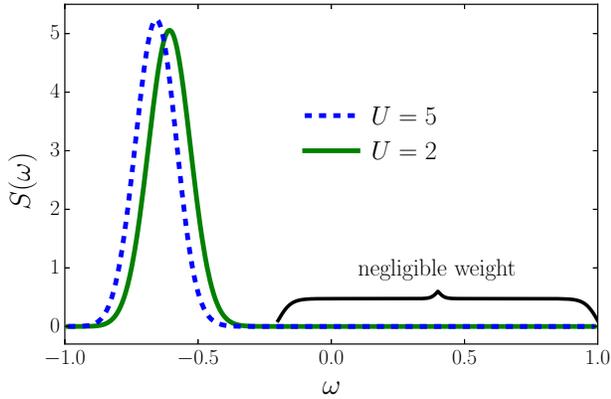}
\caption{(Color online)
The spectral decomposition $S(\omega)$ of the initial state $|\psi_0\rangle$ as defined in Eq.~\eqref{eq:SN} obtained by the procedure explained in Sec.~\ref{ptCheMPS}. The green solid line shows the spectral decomposition of $|\psi_0\rangle$ at interaction strength $U=2$, the dashed blue line shows the spectral decomposition at $U=5$. In both cases the spectral weight is located at the lower end of the spectrum and is negligible at higher energies. This allows one to determine proper projections for reduced effective bandwidths and to use the projective $t$-CheMPS method. 
}
\label{fig:spectral_decomp}
\end{figure}

\begin{figure*}[]
\includegraphics[width=6.3in]{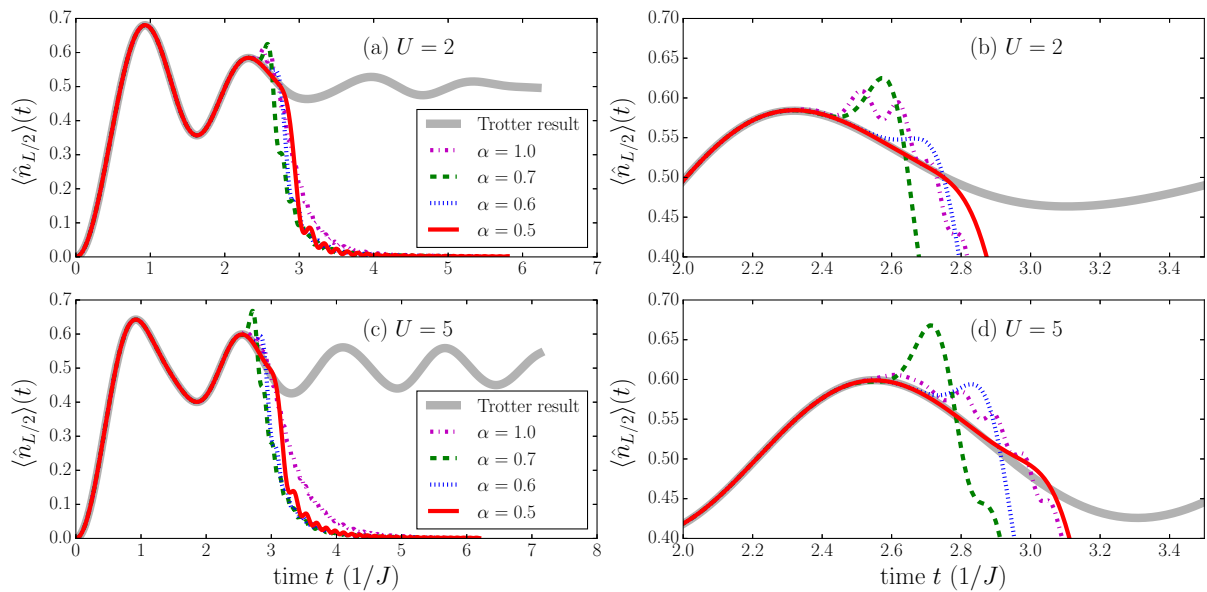}
\caption{(Color online)
Projective $t$-CheMPS results for the particle density at $L/2$ of the global quench with (a) $U=2$ and (c) $U=5$ and different projection factors $\alpha=0.5,0.6,0.7$ for an effective bandwidth $W^*=\alpha W$ as defined in Eq.~\eqref{eq:bandwith}, laid over the corresponding non-projective ($\alpha=1$) $t$-CheMPS results. Panels (b) and (d) show the same data as (a) and (c) but zoomed in to the relevant times, where errors start to diverge. Larger evolution times can be reached with the projected effective bandwidth in addition to less computational effort in the calculations and a smaller required disk space for representation of dynamics.
}
\label{fig:projective}
\end{figure*}

\subsection{Convergence and performance}
A first quantity to gauge the convergence parameters of all three methods under consideration is provided by the particle density of this global quench with $U=0$. The results shown in Fig. \ref{fig:convergence} exhibit good convergence for our purposes where the maximum on-site occupation number in $t$-DMRG is set to $\langle\hat{n}\rangle_{\text{max}}=10$ in accordance with Ref. \onlinecite{Flesch}. It is worth mentioning at this point that the underlying intention behind this work is not to contrive a method that surpasses standard methods such as Trotter time evolution and Krylov time evolution in terms of accuracy, as the latter have proven to be very precise with the right set of parameters in place. However, the goal is to investigate whether an alternative method such as $t$-CheMPS can, at the same accuracy or that within what is acceptable from an experimentally-suitable point of view, achieve larger times than those possible in Trotter time evolution or Krylov time evolution, especially that it has been demonstrated that Chebyshev polynomials can be very useful in time evolution at least outside of the context of MPS\cite{Dutta}. 

In this work, we use a $2^{\text{nd}}$-order Trotter decomposition as in previous work\cite{Honer} it has shown to be far more efficient than either $1^{\text{st}}$- or $4^{\text{th}}$-order Trotter decompositions in terms of accuracy and computational effort, respectively, while achieving approximately the same evolution times. The Krylov method we use employs an Arnoldi iteration, which is considered to be the most efficient in Krylov implementations\cite{Garcia}. In our calculations, we find that our Trotter calculations are convergent for a time step $\Delta t = 0.01/J$ and truncation or fidelity threshold\cite{White1992,Schollwoeck,White1993} of $10^{-8}$ for each time step, while Krylov and $t$-CheMPS calculations are convergent for a fidelity threshold of $10^{-5}$ for each time step. As shown in Figs. \ref{fig:convergence} and \ref{fig:denden} for the particle density and density-density correlations, respectively, Trotter decomposition is the best method when it comes to largest accessible times. The times achieved by the Krylov method (shown in Fig. \ref{fig:convergence}) match those arrived at by Flesch \emph{et al.} in Ref. \onlinecite{Flesch} for the same system. The accuracy of the $t$-CheMPS method is quite impressive and its results are actually quite exact for short times, but it can exceed neither the Krylov method nor the Trotter method in terms of largest evolution times reached.

\subsection{Projective $t$-CheMPS results}
As a further attempt at improving the results attained by the $t$-CheMPS method, we undertake the spectral decomposition in the cases of $U=2$ ($\langle\hat{n}\rangle_{\text{max}}=8$) and $U=5$ ($\langle\hat{n}\rangle_{\text{max}}=4$), the spectral functions of which are shown in Fig. \ref{fig:spectral_decomp}. The case of $U=0$ is not included as the bandwidth cannot be further reduced from its full size. Immediately, one notices that they both have nonzero weight mostly on the left half of the energy axis, \emph{i.e.} in the lower energy half of the bandwidth. In the calculations performed for this study, it has proven necessary to set $E_g^*=E_g$ to keep the Chebyshev approximation convergent, while $E_s^*\in[E_g+W/2,E_s)$. Here, $E_s$ is projected according to 

\begin{equation}\label{eq:bandwith}
E_s^*=E_g+\alpha\cdot W,
\end{equation}

\noindent where $\alpha\in(0,1]$ is the projection factor, and when $\alpha=1$, it is in fact non-projective $t$-CheMPS that is being used and energy truncation is turned off in the simulations.

Reducing the full bandwidth of the system to a reduced effective bandwidth may lead to less computational effort as one now requires fewer Chebyshev vectors in order to reach a certain maximum evolution time $t_{\text{max}}$, which is related to the expansion order $N_{\text{max}}$ by approximately\cite{Holzner} $t_{\text{max}}\approx N_{\text{max}}/a$, though, for our purposes, it is slightly smaller (see Eq.~\eqref{eq:fabian}) in order to achieve a desired precision (Sec. \ref{timeconstraint}). Since $a$ scales proportionally with the reduction in the full bandwidth upon projection, one need only achieve the same number of vectors in projective $t$-CheMPS as in non-projective $t$-CheMPS to facilitate a maximum evolution time bigger by that same factor of reduction in the full bandwidth. However, in projective $t$-CheMPS a new function enters into the computation, namely that of energy truncation. In our numerical simulations, the projective $t$-CheMPS method at any factor of reduction is unable to calculate up to the same order of expansion as the non-projective $t$-CheMPS method due to the additional computational effort of energy truncation, but, nevertheless, for certain projections, evolution times bigger than those achieved in non-projective $t$-CheMPS are reached as shown in Fig. \ref{fig:projective} for both cases $U=2$ and $U=5$. The best result is attained for both $U$ values at $\alpha=0.5$, the most stringent projection factor used that did not lead to divergences, where an improvement of $20\%$ ($12\%$) is achieved for $U=2$ ($U=5$) in terms of largest accessible evolution times. The corresponding converged Trotter results are overlaid for reference. It can be seen that even though projective $t$-CheMPS does indeed reach greater evolution times than its non-projective counterpart, it still does not improve over the Trotter or Krylov methods. It is also worth noting here that projective $t$-CheMPS, despite even sometimes significant reductions in the full bandwidth of the system, still offers exact results for short times. 

Though one may be tempted to think that the projective $t$-CheMPS method must achieve longer times the more one projects (\emph{i.e.}, the smaller $\alpha$ is), this is not the case in reality as then more computational effort is required by the energy-truncation module that at some point it simply cannot handle all the required energy projections when $\alpha\ll 1$, and in fact this renders the maximum evolution time reachable smaller than that in the non-projective $t$-CheMPS method or it may outright lead to divergences\cite{Holzner}. On the other extreme, if $\alpha\lesssim 1$, then $W^*\lesssim W$, and thus the computational effort is almost the same as in the non-projective $t$-CheMPS method with the added cost of energy truncation, which leads to evolution times shorter than those attained by non-projective $t$-CheMPS. Thus, one has to choose $\alpha$ in a manner where energy truncation is not pushed to its limits and while at the same time $W^*$ is nontrivially smaller than $W$.

\begin{figure*}[]
\includegraphics[width=6.3in]{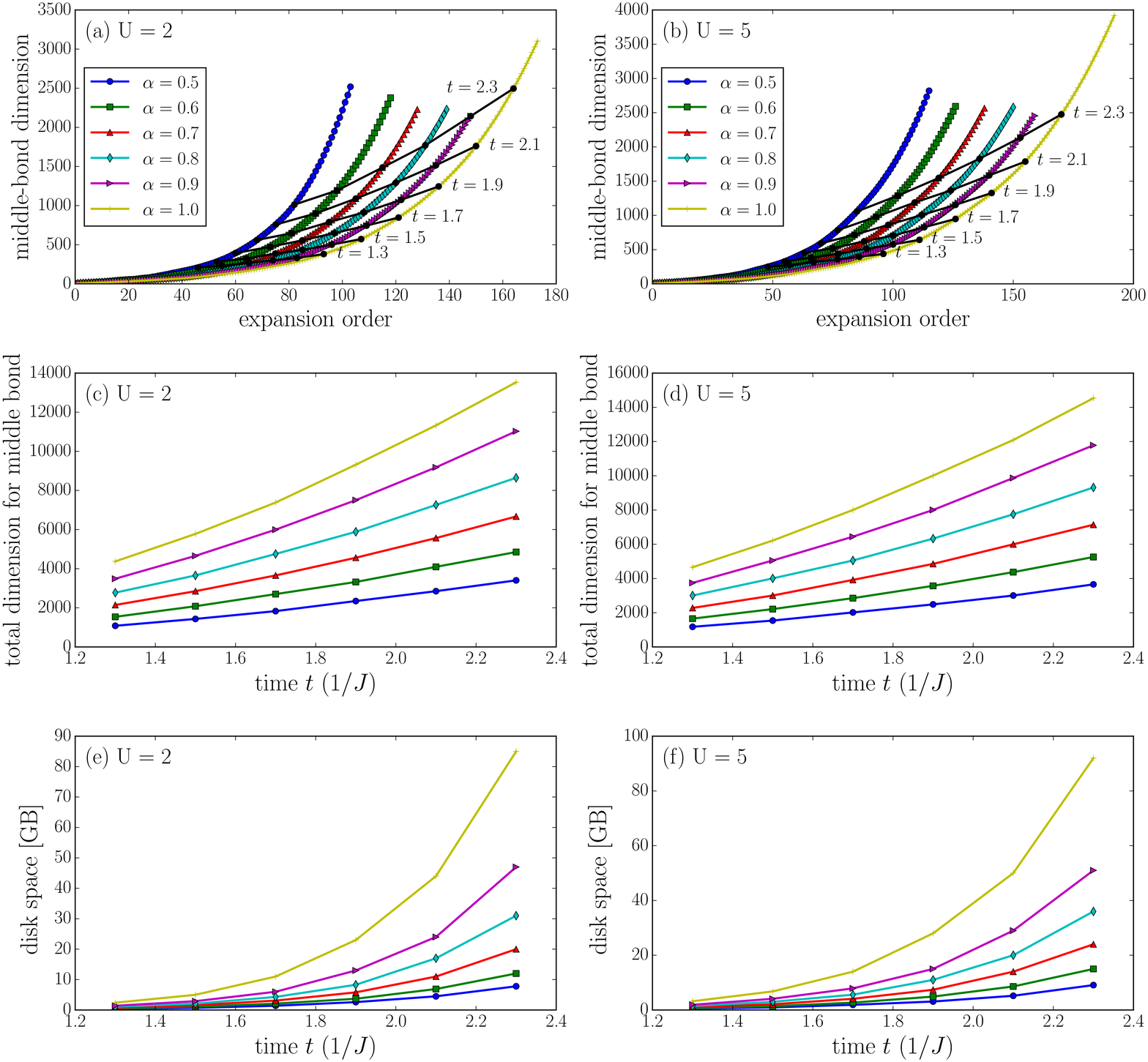}
\caption{(Color online)
The matrix dimension at the central bond for the Chebyshev vectors due to the projective ($\alpha<1$) and non-projective ($\alpha=1$) $t$-CheMPS methods for $U=2$ (left column) and $U=5$ (right column). In (a) and (b), one notices that at any expansion index $n$, the corresponding Chebyshev vector $|t_n\rangle$ carries a larger central-bond dimension the smaller $\alpha$ is (\emph{i.e.}, the more reduced the effective bandwidth is), as at the corresponding expansion order $n+1$, $t$-CheMPS represents longer dynamics the smaller $\alpha$ is. However, in (c) and (d), one notices that the sum of the matrix dimensions at the central bond for the Chebyshev vectors leading up to a certain time $t$ is far smaller the lower the value of $\alpha$, indicating less computational effort upon greater reduction in the bandwidth. Note how the projective $t$-CheMPS vectors cannot reach the maximum matrix dimension non-projective $t$-CheMPS vectors have, and this is due to the additional computational effort of energy truncation necessary in the projective $t$-CheMPS method but nonexistent in its non-projective counterpart. Additionally, (e) and (f) show significant conservation of disk space in projective $t$-CheMPS for any evolution time $t$, where the smaller $\alpha$ is, the less disk space is required for storing the Chebyshev vectors that are necessary to represent dynamics up to $t$. Note that the selected times are indicated via temporal isolines in (a) and (b).
}
\label{fig:states}
\end{figure*}

\subsection{Middle-bond dimension and vector size}
To avoid any confusion, we remind the reader here that the expansion \emph{order} $N$ is the \emph{total number} of Chebyshev vectors, and each of the latter is indicated by an expansion \emph{index} $n$ that goes from $0$ for the first vector to $N-1$ for the highest-index vector.

Intuitively, the projective $t$-CheMPS vectors are expected to comprise of higher bond dimensions at the same expansion order than their non-projective $t$-CheMPS counterparts as exhibited in Fig. \ref{fig:states}(a) and (b) for $U=2$ and $U=5$, respectively, at the middle or central bond. This can be attributed to the fact that for the same time $t$ attained in both methods, expansion order $N^*$ achieved by projective $t$-CheMPS for faithful representation of the system dynamics at this time is related to the corresponding expansion order $N$ attained by non-projective $t$-CheMPS through $N^*\approx\alpha\cdot N<N$. Hence, the vector of a certain expansion index carries more information when generated by projective rather than non-projective $t$-CheMPS, because the generated $N^*$ Chebyshev vectors in projective $t$-CheMPS still, assuming convergence, must carry the same information about the system as the $N$ ($\approx N^*/\alpha>N^*)$ Chebyshev vectors in non-projective $t$-CheMPS do. However, it can be seen in Fig. 6(a) and (b) that the projective $t$-CheMPS vectors do not reach the maximum central-bond dimensions that occur for the non-projective $t$-CheMPS vectors. In principle, one may expect that all the $t$-CheMPS vectors, regardless of the value of $\alpha$ ought to reach the same maximal matrix dimensions. This is only true, however, if the workings of these calculations are the same, but this is not the case because in projective ($\alpha<1$) $t$-CheMPS an additional computation effort is needed, that of energy truncation, which does not occur in non-projective ($\alpha=1$) $t$-CheMPS.

In addition to obtaining fewer vectors required to arrive at an evolution time $t$ in projective $t$-CheMPS, one finds that these vectors are in fact smaller in size the bigger the reduction in bandwidth, \emph{i.e.}, the smaller $\alpha$, is. In Fig. 6(a) and (b), each temporal isoline is constructed for a properly selected evolution time $t$ that is appropriately matched to its corresponding expansion orders $N_{\alpha}$ for the different $\alpha$ values based on the criterion in Eq.~\eqref{eq:fabian} (one may equally well use the more relaxed criterion of $N_{\alpha}\approx\alpha Wt/(2-\epsilon_t)$, which our calculations show is also adequate for $U=2$ and $U=5$). These temporal isolines indicate that at a time $t$, the corresponding projective and non-projective $t$-CheMPS maximum-expansion-index vectors $|t_{N^*-1}\rangle$ and $|t_{N-1}\rangle$, respectively, are such that the latter has larger matrix dimensions than the former, and this becomes more pronounced the larger $t$ is. It is interesting to also look at the total sum of matrix dimensions at the central bond of the Chebyshev vectors involved in arriving at a time $t$ in $t$-CheMPS for different values of $\alpha$. If $D_n=D(|t_n\rangle)$ indicates the matrix dimension at the central bond of $|t_n\rangle$, then $\sum_{n=0}^{N_{\alpha}-1}D_n$ would be a good measure of the computational effort required to reach a time $t$ based on Eq.~\eqref{eq:fabian} in (non-)projective $t$-CheMPS for some value of $\alpha$. This measure not only incorporates the maximal matrix dimension attained by the highest-index Chebyshev vector required to reach an evolution time $t$, but it also accounts for how many vectors are required to reach $t$, and this number varies depending on the value of $\alpha$. This measure is depicted in Fig. 6(c) and (d) for $U=2$ and $U=5$, respectively, where one can conclude that the more reduced the effective bandwidth is (the smaller $\alpha$ is), the smaller is the computational effort required to reach a certain evolution time $t$. For the longest common time arrived at by all calculations ($t\approx 2.3/J$), there is a factor of roughly $4$ with regards to mitigation of computational effort from $\alpha=1$ to $\alpha=0.5$. 

Moreover, at a certain evolution time $t$ corresponding to a set of expansion orders $N_{\alpha}$ for the different $\alpha$-valued $t$-CheMPS calculations, one finds that the highest-index Chebyshev vector $|t_{N_{\alpha}-1}\rangle$ occupies less disk space the smaller $\alpha$ is. If $d_n=d(|t_n\rangle)$ is the disk space occupied by Chebyshev vector $|t_n\rangle$, then $\sum_{n=0}^{N_{\alpha}-1}d_n$ is the total disk space needed to house those Chebyshev vectors required to arrive at the dynamics up to time $t$ corresponding to $N_{\alpha}$ as per Eq.~\eqref{eq:fabian}. This is presented in Fig. 6(e) and (f) for $U=2$ and $U=5$, respectively, where it can be seen that the greater the projection (or the smaller $\alpha$ is), the more reduction one obtains in total disk space. In fact, at $t=2.3/J$, the reduction is more than an order of magnitude from $\alpha=1$ to $\alpha=0.5$. Therefore, upon projection, one obtains fewer Chebyshev vectors that as a whole are also smaller in size while representing the same dynamics, which indicates data compression.

\begin{figure}[]
\includegraphics[scale=0.45]{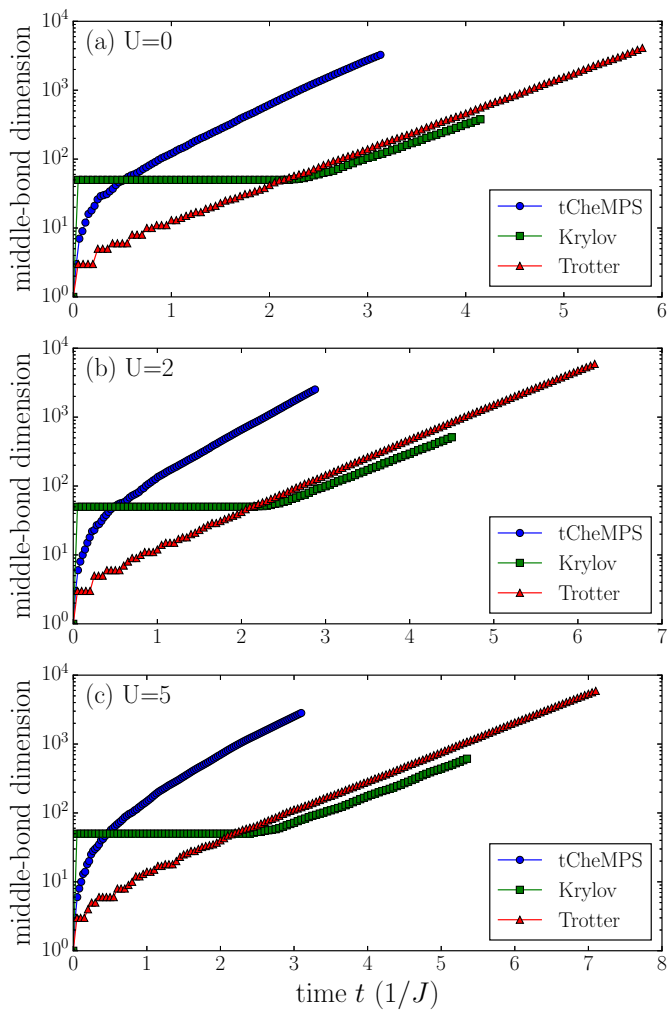}
\caption{(Color online)
The matrix dimension at the central bond of the bosonic chain using the $t$-CheMPS, Krylov, and Trotter methods for interaction strengths (a) $U=0$, (b) $U=2$, and (c) $U=5$. The matrix dimension for $t$-CheMPS increases very quickly and greatly exceeds its counterparts in the Krylov and Trotter methods for common evolution times irrespective of the interaction strength. This prohibits the method from achieving much longer times.
}
\label{fig:comparison}
\end{figure}

\subsection{Comparison with other methods}
It is interesting to produce a quantitative comparison of the $t$-CheMPS method with the Krylov and Trotter methods in the time domain. One can consider the matrix dimension at the central bond required to faithfully represent the dynamics over the evolution times. One can again here represent the central-bond total matrix dimension for $t$-CheMPS at a time $t$ as $\sum_{n=0}^{N_{\alpha}-1}D_n$, where $D_n$ is the matrix dimension at the central bond of $|t_n\rangle$, and $N_{\alpha}$ and $t$ are related as per Eq.~\eqref{eq:fabian}, as is done in Fig. \ref{fig:states}, but this representation would not be fair as it encompasses all the Chebyshev vectors required to construct the wavefunction $|\psi(t)\rangle$, the construction of which, unlike in the Trotter or Krylov methods, is never undertaken in the $t$-CheMPS method (see Sec.~\ref{timeconstraint}) . Thus, even though this representation is proper in Fig. \ref{fig:states}(c) and (d) as it involves a comparison between the different bandwidth reductions in the $t$-CheMPS method through covering the number of Chebyshev vectors involved in reaching an evolution time $t$, for comparison with the Trotter and Krylov results, $D_{N_{\alpha}-1}$ is the proper quantity to look at, because $D_{N_{\alpha}-1}$ is the largest matrix dimension of the central bond attained by any of the Chebyshev vectors required for faithful representation of the dynamics up to evolution time $t$. For this comparison, the $\alpha=0.5$ projective $t$-CheMPS result for $U=2$ and $U=5$ is chosen as it performs best compared to other $t$-CheMPS approaches at those interaction strengths, while the non-projective $t$-CheMPS result is used for $U=0$ as there no projection is possible. The comparison is displayed in Fig. \ref{fig:comparison}, where it can be noted that the central-bond matrix dimension required in the $t$-CheMPS method to represent the dynamics up to a common evolution time $t$ is much greater than that in the Krylov or Trotter methods regardless of what the interaction strength $U$ is. This is in agreement with Ref.~\onlinecite{Wolf2}, particularly in the case where the rescaled Hamiltonian $\hat{H}'$ is simply a factor of the original Hamiltonian $\hat{H}$, which is the case in this study when $U=0$, depicted in Fig. \ref{fig:comparison}(a). At this interaction strength, the skystate and groundstate energies are equal in magnitude but of opposite sign, rendering $b=0$. This leads to $\hat{H}'=\hat{H}/a$, which is the condition proven in Ref.~\onlinecite{Wolf2} to assert that then the Chebyshev vectors are equivalent to time-evolved wavefunctions for a proper time step in the Krylov or Trotter methods\cite{Wolf2}.

\subsection{Matrix moments}
\begin{figure}[]
\includegraphics[width=1.\linewidth]{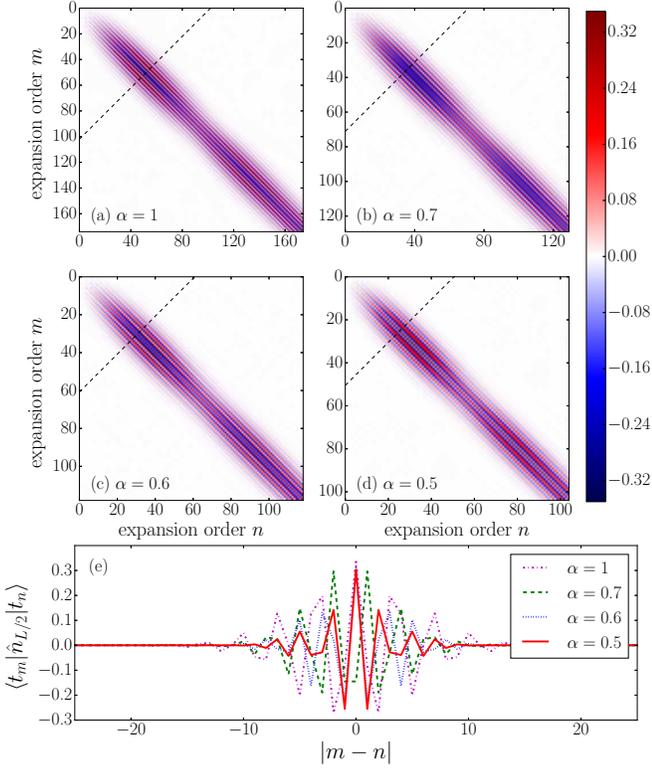}
\caption{(Color online)
The behavior of the Chebyshev moments $\langle t_m|\hat{n}_{L/2}|t_n\rangle$ for (a) non-projective ($\alpha=1$) and projective $t$-CheMPS for (b) $\alpha=0.7$, (c) $\alpha=0.6$, and (d) $\alpha=0.5$. It can be seen how the bulk of the information lies where $|m-n|<15$ and biggest around $m\approx n$. In (e) the cross sections of these moments along the back diagonal $\{(N_{\alpha},0),(0,N_{\alpha})\}$ where $N(\alpha)=100\alpha$ are shown, displaying rapid decay around $|m-n|=0$ for all $\alpha$ values.
}
\label{fig:cheb_matrix2}
\end{figure}

Finally, in Fig. \ref{fig:cheb_matrix2}, we take a closer look at the behavior of the Chebyshev moments $\langle t_m|\hat{n}_{L/2}|t_n\rangle$. In particular, we observe that these moments carry the greatest weight along the back diagonal (\textbackslash). This is to be expected as these Chebyshev moments involve two states $|t_n\rangle$ and $|t_m\rangle$ that have little overlap, since, if $m>n$, $|t_m\rangle$ is arrived at by consecutively applying $\hat{H}'$ $m-n$ times onto $|t_n\rangle$, and as the latter is not an eigenstate of $\hat{H}'$, this renders the two vectors with little overlap the bigger $|m-n|$ is. Moreover, the cross sections of these moments along the dotted black lines in Fig. \ref{fig:cheb_matrix2}(a)-(d), corresponding to some expansion order, say $N(\alpha)=100\alpha$, exhibit a decaying behavior around $|m-n|=0$ that is zero for large $|m-n|$. These cross sections for the different $\alpha$ values are depicted in Fig. \ref{fig:cheb_matrix2}(e). As mentioned previously, this validates the constraint for the maximum evolution time $t_{\text{max}}$ that $\sum_{n=N}^{\infty}\phi_{n}^{*}(t)\phi_{n}(t) < 10^{-3}$ while neglecting off-diagonal terms as indeed one can see that the Chebyshev moments in Fig. \ref{fig:cheb_matrix2} carry nontrivial weight mostly for quite small values of $|m-n|$, thereby making this constraint sufficient.

\section{Conclusion}\label{conclusion}
A new method, $t$-CheMPS, based on the Chebyshev expansion in the time domain in the context of MPS has been presented for calculating the time evolution of quantum many-body systems, including global quenches. Using a test system of importance in the field of quantum many-body physics, we demonstrate that $t$-CheMPS arrives at exact solutions of a given observable for short times, but does not exceed the largest times accessible by standard time-evolution methods such as the Trotter decomposition and the Krylov approximation. Furthermore, a projective version of the method, projective $t$-CheMPS, based on spectral decomposition and system-bandwidth reduction is introduced that improves on the largest evolution times accessible while significantly easing computational effort and greatly reducing disk space for the same dynamics. Moreover, we find again that Trotter expansion is still the favorable method with regards to largest accessible evolution times.

\section{Acknowledgments}
 The authors are grateful to Thomas Barthel (Duke University), Alex Wolf, Andreas Holzner, Andreas Weichselbaum, and Jan von Delft (all of LMU Physik) for fruitful discussions, and to Ulrich Schollw\"ock (LMU Physik) for a thorough reading of and subsequent valuable comments on the manuscript.  J.C.H. acknowledges support through the FP7/Marie-Curie grant $321918$ and DFG FOR $801$. I.P.M. acknowledges the support from the Australian Research Council Centre of Excellence for Engineered Quantum Systems, CE110001013, and the Future Fellowships scheme, FT100100515..
 
\appendix*
\section{Spectral decomposition of the time-evolved wavefunction $|\psi(t>0)\rangle$}
Reminding the reader that here for ease of notation, as in Sec.~\ref{ptCheMPS}, $\hat{H}$ is taken to be the rescaled Hamiltonian with bandwidth $W=E_s-E_g$, and $\omega\in[-1,1]$, we proceed with first remarking that the spectral decomposition of the time-evolved state $|\psi(t)\rangle$ is equal to that of the initial state $|\psi_0\rangle$:

\begin{align}
\langle\psi(t)|&\delta(\omega-\hat{H})|\psi(t)\rangle \nonumber \\ \nonumber
&=\langle\psi_0|\hat{U}^{\dagger}(t)\delta(\omega-\hat{H})\hat{U}(t)|\psi_0\rangle \\ 
&=\langle\psi_0|\delta(\omega-\hat{H})|\psi_0\rangle,
\end{align}

\noindent where the time-evolution operator $\hat{U}(t)$ commutes with $\delta(\omega-\hat{H})$. Thus, it is a good check of the validity and convergence of the Chebyshev vectors to ascertain that the spectral decomposition is the same at any time $t$ when calculated by the corresponding Chebyshev moments. Furthermore, this may also be employed as an alternate way to Eq.~\eqref{eq:fabian} to determine how many Chebyshev vectors one would need to faithfully represent the physics at an evolution time $t$, using the error with respect to $S(\omega)$ in Eq. (\ref{eq:SN}) as a gauge. As such, we provide here a derivation that allows one to calculate the spectral decomposition of the time-evolved wavefunction $|\psi(t)\rangle$ from the corresponding Chebyshev moments.

In the $t$-CheMPS method, one can represent the wavefunction $|\psi(t)\rangle=\hat{U}(t) |\psi_0\rangle$ as 

\begin{equation} 
\label{eq:psiN} |\psi(t)\rangle = e^{-i(E_g+aW')t}\sum\limits_{n=0}^{N-1}\phi_n(t)|t_n\rangle
\end{equation}

\noindent as per Eq.~\eqref{eq:obs1}. To reach a specified evolution time $t$, we need to use an expansion order of $N_t\approx tW/2$ or, more stringently, as specified by Eq.~\eqref{eq:fabian}.

Now, to calculate the spectral function $S_t(\omega)$ for the wavefunction $|\psi(t)\rangle$, with a specified spectral resolution $\Delta$, we can proceed as follows:

\begin{align} \label{eq:StN} \nonumber S_t(\omega) &= \langle\psi(t)|\delta_{N_\Delta}(\omega-\hat{H})|\psi(t)\rangle \\ \nonumber
&= \sum\limits_{n,n'=0}^{N_t-1}\phi^{*}_{n'}(t)\phi_n(t)\langle t_{n'}|\delta_{N_\Delta}(\omega-\hat{H})|t_n\rangle \\
&= \frac{1}{\pi\sqrt{1-\omega^2}} \sum\limits_{n,n'=0}^{N_t-1}\sum\limits_{n''=0}^{N_\Delta-1}\phi^\ast_{n'}(t)\phi_n(t)\phi_{n''}T_{n''}(\omega)\mu_{n''}^{n'n},
\end{align}

\noindent where $\mu_{n''}^{n'n}=\langle t_{n'}|T_{n''}(\hat{H})|t_n\rangle$.

Note that we need to use different upper limits on the sums on $n$ and $n'$ than on the sum on $n''$, in order to reach a specified time $t$ with a specified spectral resolution $\Delta$. To evaluate the moments arising here, we recall the following identity:

\begin{equation} \label{eq:ChebyIdent} T_{n_1}(\hat{H})T_{n_2}(\hat{H}) = \frac{1}{2}T_{n_1+n_2}(\hat{H})+\frac{1}{2}T_{|n_1-n_2|}(\hat{H}). \end{equation}

\noindent It is advisable to use it in such a way that the order of the polynomials that arise remain as small as possible. Thus, for the case that $n<n'$, we proceed as follows (with $n_1=n''$ and $n_2=n$):

\begin{equation} \label{eq:ChebyIdent1} T_{n''}(\hat{H})T_{n}(\hat{H}) = \frac{1}{2}T_{n''+n}(\hat{H})+\frac{1}{2}T_{|n''-n|}(\hat{H}), \end{equation}

\noindent which leads to

\begin{align} \label{eq:mu} \mu_{n''}^{n'n} \nonumber & =\langle t_{n'}|T_{n''}(\hat{H})|t_n\rangle = \langle t_{n'}|T_{n''}(\hat{H})T_{n}(\hat{H})|t_0\rangle \\ \nonumber
& = \frac{1}{2}\langle t_{n'}|T_{n+n''}(\hat{H})|t_0\rangle+\frac{1}{2}\langle t_{n'}|T_{|n-n''|}(\hat{H})|t_0\rangle \\
&= \frac{1}{2} \langle t_{n'}|t_{n+n''}\rangle+\frac{1}{2}\langle t_{n'}|t_{|n-n''|}\rangle.
\end{align}

\noindent For the case that $n'<n$, we proceed analogously, but with $n_1=n'$ and $n_2=n''$:

\begin{equation} \label{eq:ChebyIdent2} T_{n'}(\hat{H})T_{n''}(\hat{H}) = \frac{1}{2}T_{n'+n''}(\hat{H})+\frac{1}{2}T_{|n'-n''|}(\hat{H}), \end{equation}

\noindent which in turn leads to

\begin{align} \label{eq:mu1} \mu_{n''}^{n'n} \nonumber & = \langle t_{n'}|T_{n''}(\hat{H})|t_n\rangle = \langle t_0|T_{n'}(\hat{H})T_{n''}(\hat{H})|t_n\rangle \\ \nonumber
& =  \frac{1}{2}\langle t_0|T_{n'+n''}(\hat{H})|t_n\rangle+\frac{1}{2}\langle t_0|T_{|n'-n''|}(\hat{H})|t_n\rangle \\
&= \frac{1}{2} \langle t_{n'+n''}|t_n\rangle+\frac{1}{2}\langle t_{|n'-n''|}|t_n\rangle.
\end{align}

\noindent Thus, we conclude that in order to calculate $S_t(\omega)$ with a specified resolution of $\Delta$ up to a specified time $t$, we need all Chebyshev vectors up to order $N_\Delta+N_t$.


\end{document}